\documentclass[%
 reprint,
preprintnumbers,
 amsmath,amssymb,
 aps,
]{revtex4-2}

\usepackage{graphicx}
\usepackage{dcolumn}
\usepackage{bm}
\usepackage{xcolor}

\newcommand{\Tr}{\mathrm{Tr\,}}
\newcommand{\floor}[1]{\left\lfloor #1 \right\rfloor}
\newcommand{\be}{\begin{equation}}
\newcommand{\ee}{\end{equation}}

\begin{document}

\preprint{CERN-TH-2026-125}

\title{BPS Non-Renormalization in the BMN Matrix Model}

\author{Sean Colin-Ellerin}
\email{sean.julian.colin-ellerin@cern.ch}
\author{Kyriakos Papadodimas}
\email{kyriakos.papadodimas@cern.ch}
\affiliation{Theoretical Physics Department, CERN, CH-1211 Geneva 23, Switzerland\\}

\date{\today}

\begin{abstract}

We show in the $(0+1)$-dimensional Berenstein-Maldacena-Nastase matrix model, dual to M-theory on a pp-wave background, that the coupling can be changed between any two finite, non-zero values using a special class of deformations, known as conjugation deformations. Importantly, we prove that they preserve normalizability of the states. This implies that BPS states in the model cannot lift as the couplings are varied, and hence their (unsigned) number cannot change, except at the free point and Banks-Fischler-Shenker-Susskind point. 

\end{abstract}

\maketitle


\section{Introduction}
\label{sec:intro}
The problem of finding the supersymmetric ground states of an interacting supersymmetric quantum theory in a fixed charge sector, or simply counting them, is notoriously difficult. Even given knowledge of them at some value of the couplings, e.g. the free point, the energy of many of these states will generically acquire quantum corrections as the couplings are varied and cease to be ground states, unless prevented from doing so by representation theory. Such states, known as BPS states, are particularly interesting in holographic quantum systems because, at strong coupling, they comprise the microstates of supersymmetric black holes in the dual gravity system. In this work, we will show that in the (0+1)-dimensional Berenstein-Maldacena-Nastase (BMN) matrix model \cite{Berenstein:2002jq} dual to M-theory in a pp-wave background, the number of BPS states cannot change as the coupling is varied between any two finite, non-zero values so it must be the same at weak and strong coupling. This is a stronger statement than the Witten index, which only gives a signed count of BPS states. The supergravity dual is known to have non-supersymmetric black holes \cite{Costa:2014wya}, and there is evidence from index computations in the matrix model that there should also exist supersymmetric black holes \cite{Chang:2024lkw, Zigdon:2025icw, Chang:2026mui}.

The BMN matrix model is a supersymmetric $(0+1)$-dimensional quantum mechanical model with $16$ supercharges consisting of $9$ $N \times N$ bosonic Hermitian matrices $X_{A}$ and $16$ real fermions $\psi_{\mathfrak{I}}$, as well as a background gauge field $A_{0}$. The theory has an $SO(3) \times SO(6)$ global symmetry so we break the bosons into $3+6$ matrices $X_{i}$ $(i=1,2,3)$ and $X_{a}$ $(a=4,\ldots,9)$ and repackage the fermions as complex $\psi_{\alpha I}$ which transform in the fundamental representation $({\bf 2},{\bf 4})$ of $SU(2) \times SU(4)$ with corresponding labels $\alpha$ and $I$. 

The Hamiltonian of the model is
\begin{align}
H &=  \Tr \left( \frac{1}{2} \Pi_A^2 - \frac{g^{2}}{4} [X_A, X_B]^2 
- \frac{g}{2} \Psi^\top \gamma^A [X_A, \Psi] \right) \nonumber
\\ & +\frac{1}{2} \Tr 
\Big({\left(\frac{\mu}{ 
3}\right)^2 } X_iX^{i} + \left(\frac{\mu}{6}\right)^2 X_aX^{a} + 
i \frac{\mu}{4} \Psi^\top \gamma^{123} \Psi \nonumber
\\  & + i \frac{2g\mu}{3} 
\epsilon^{ijk} X_i X_j X_k \Big)\,,   
\end{align}
where $\Pi_{A}$ is the conjugate momentum to $X_{A}$, $\Psi = \left(\begin{smallmatrix} \psi_{\alpha I} \\ \epsilon_{\alpha\beta}\psi^{\dagger\beta I}\end{smallmatrix}\right)$, and we have set the Planck length $\ell_{P}=1$ \footnote{This Hamiltonian is related to the standard one in eq. (1) of \cite{Dasgupta:2002ru} by $g=R^{\frac{3}{2}}$ and the redefinition $\Pi^{\mathrm{us}} = g^{\frac{1}{3}}\Pi^{\mathrm{them}}$ and $X^{\mathrm{us}} = g^{-\frac{1}{3}}X^{\mathrm{them}}$. Our choice gives a canonically normalized kinetic term. We use the same conventions as them for spinors, gamma matrices, etc.}. The model has a single dimensionless coupling $\lambda = g^{\frac{2}{3}}/\mu$ so we can set $g=1$, leaving $\mu$ as the only coupling, and restore $g$ by dimensional analysis. The Hamiltonian has an $SU(N)$ gauge symmetry under which the operators transform in the adjoint representation, which is imposed by the background gauge field $A_{0}$ \footnote{There is a trivial $U(1)$ sector of the model coming from the trace mode, which is always free, and has an extra 16 supercharges associated to it. The non-trivial, interacting $SU(N)$ sector consists of traceless, Hermitian matrices. We will restrict attention to the non-trivial part of the theory.}.

The classical vacua of the model are the fuzzy spheres $X_{a}=0$ and $X_{i}=\frac{\mu}{3g}J_{i}$ where $J_{i}$ is the $i$th generator of an $N$-dimensional representation of $SU(2)$, which are in one-to-one correspondence with partitions $\mathcal{P}=(N_{1},\ldots,N_{k})$ of $N$. By analyzing the Gauss Law constraints, one finds that the physical states in vacuum sector $\mathcal{P}$ must be built from $U(N_{1}) \times \ldots \times U(N_{k})$-invariant operators.

For any $\mu > 0$, the model has a discrete spectrum and becomes a set of decoupled harmonic oscillators at $\mu=\infty$---the free point---with each vacuum its own superselection sector in this limit. On the other hand, at $\mu=0$, it becomes the Banks-Fischler-Shenker-Susskind (BFSS) model \cite{Banks:1996vh}, which is known to have a unique normalizable, supersymmetric ground state above which the spectrum is continuous \cite{DEWIT1989135, Sethi:1997pa, Porrati:1997ej, Moore:1998et}. The M-theory limit is $R=g^{\frac{2}{3}},N \to \infty$ with $p^{+}=N/R$ fixed.  For a detailed analysis of the model, we refer the reader to \cite{Dasgupta:2002hx}.

The BMN model has a rich spectrum of BPS states analyzed in \cite{Dasgupta:2002ru, Kim:2002zg}, which will be discussed in more detail in Section \ref{sec:BPS}. The only zero-charge BPS states are the vacua, which are fully BPS (annihilated by all of the supercharges), and the excited, charged BPS states are 1/2-, 3/8- 1/4-, or 1/8-BPS \footnote{Note that in the literature sometimes the 16 supercharges from the trivial $U(1)$ sector are also included in the counting, so the designation of BPS is $1/2$ of ours, e.g. the vacuum would be 1/2-BPS in those conventions while it is fully BPS in ours.}. The BPS states are not known at generic coupling and, for most of those present in the free theory, it is not known whether they remain BPS in the interacting theory. 

We will demonstrate that the number of them at finite, non-zero coupling is constant as a function of the coupling using the following argument due to Witten \cite{WITTEN1982253}. The BPS subspace preserving any two real supercharges is isomorphic to the corresponding supercharge cohomology. If one can find a similarity transformation acting on the supercharges that changes the couplings, known as a conjugation deformation, then the cohomologies at the different values of couplings will be isomorphic, and hence the number of BPS states cannot change. The main result of this work is to construct a single similarity transformation that changes the coupling in all of the supercharges in the BMN model to any value, except the exceptional points consisting of the BFSS point $(\mu=0)$ and free point $(\mu=\infty)$. Crucially, we show that this transformation preserves normalizability of the states. We then apply this result to understand some features of the BPS spectrum and the chiral ring.

\textbf{Note added:} As this work was in progress, we learned of independent work on this problem by Chi-Ming Chang, Zhengyuan Du, Sarthak Duary, Kangning Liu, and Yi-Xiao Tao, and thus have coordinated submission.

\section{Conjugation Deformations}
\label{sec:conjdef}

We will first review the supersymmetry algebra of the BMN model and the basics of conjugation deformations, before presenting the deformation operator for BMN along with its range of validity. 

\subsection{Supersymmetry algebra}

The $16$ real supercharges organize into $8$ complex supercharges $Q_{\alpha I}$ and their conjugates $Q_{\alpha I}^{\dagger}$, expressed in terms of the elementary operators as
\begin{align}
\begin{split}
\label{eqn:supercharges}
Q_{\alpha I} &= \Tr \bigg( (\Pi_a - i \frac{\mu}{6} X_a)
{\sf g}^a_{IJ} 
\epsilon_{\alpha \beta} \psi^{\dagger J \beta} - (\Pi_i + i \frac{\mu}{3} X_i)
 \sigma^i_\alpha {}^\beta \psi_{I \beta} 
\\ &+ \frac{g}{2} [X_i, X_j] \epsilon^{ijk} \sigma^k_\alpha {}^\beta 
\psi_{I \beta} - \frac{ig}{2} [X_a, X_b] ({\sf g}^{ab})_I^J \psi_{J \alpha} 
\\  &+ig 
[X_i, X_a] \sigma^i_\alpha {}^\lambda {\sf g}^a_{IJ} \epsilon_{\lambda \beta} \psi^{\dagger J 
\beta} \bigg)\,,
\end{split}
\end{align}
which transform in the $(\mathbf{2},\mathbf{4})$ of the $SU(2) \times SU(4)$ $R$-symmetry with corresponding indices $\alpha=\pm$ and $I=1,\ldots,4$ . The intertwiners ${\sf g}^a_{IJ}$ and $\sigma^k_\alpha {}^\beta$ form the blocks of the $SO(9)$ gamma matrices $\gamma^a = \left( \begin{smallmatrix} 0 & 1 \times {\sf g}^a \\ 1 \times ({\sf 
g}^a)^\dagger & 0 \end{smallmatrix} \right)$ and $\gamma^i = \left( \begin{smallmatrix} -\sigma^i \times 
1 & 0  \\ 0 & \sigma^i \times 1 \end{smallmatrix} \right)$.

These supercharges satisfy the algebra
\begin{equation}
\label{eqn:SUSYalg}
\{Q^{\dagger \alpha I},Q_{\beta J} \} = 2 \delta^I_J \delta^\alpha_\beta H   
+ \frac{\mu}{3} \epsilon^{ijk} \sigma^k_{\beta} {}^\alpha \delta^I_J
M^{ij} - \frac{i\mu}{6}  \delta^\alpha_\beta ({\sf g}^{ab})_J {}^I
M^{ab}
\end{equation}
where $M^{ij}$ and $M^{ab}$ are the $SO(3)$ and $SO(6)$ generators, respectively. The supercharges do not commute with the Hamiltonian because the supersymmetry transformations are time-dependent, instead they raise and lower the energy by $\mu/12$, viz.
\begin{equation}
    [H,Q_{\alpha I}] = \frac{\mu}{12}Q_{\alpha I}, \qquad [H,Q_{\alpha I}^{\dagger}] = -\frac{\mu}{12}Q_{\alpha I}^{\dagger}\,.
\end{equation}
Hence, the states in a supermultiplet have different energies.

Given a supercharge $Q_{\alpha I}$, we say a state $|\Phi\rangle$ is BPS with respect to this supercharge if $Q_{\alpha I}|\Phi\rangle=Q_{\alpha I}^{\dagger}|\Phi\rangle=0$. If this holds for $k$ out of the $8$ pairs of supercharges, then the state is $k/8$-BPS. Since the charges of the global symmetries appear on the righthand side of eq. \eqref{eqn:SUSYalg}, we see that a BPS state generally carries non-trivial charges.
\subsection{Conjugation deformations}

Let us now review a special class of deformations, known as conjugation deformations, and the argument for why they cannot change the number of BPS states \cite{WITTEN1982253}. 

Choose a supercharge $Q_{\alpha I}$. By standard Hodge theory, the Hilbert space of BPS states $\mathbb{H}_{\mathbf{BPS}}(Q_{\alpha I})$ with respect to $Q_{\alpha I}$ is isomorphic to the $Q_{\alpha I}$-cohomology $H^{\bullet}(Q_{\alpha I})$ of states that are $Q_{\alpha I}$-closed modulo $Q_{\alpha I}$-exact states. Consider an invertible transformation $M$ (not necessarily unitary). Acting on the supercharge via
\begin{equation}
Q_{\alpha I} \to \widetilde{Q}_{\alpha I} = M^{-1}Q_{\alpha I}M\,,
\end{equation}
this gives an isomorphism of the cohomologies $H^{\bullet}(Q_{\alpha I}) \cong H^{\bullet}(\widetilde{Q}_{\alpha I})$ with the explicit map between $Q_{\alpha I}$-cohomology representatives given by $\phi \mapsto M^{-1}\phi$.

Therefore, if we can find an $M(\mu_{1},\mu_{0})$ that changes the value of the coupling in the supercharge $Q_{\alpha I}(\mu_{0}) \to Q_{\alpha I}(\mu_{1})$, then the BPS Hilbert spaces must be isomorphic. An important subtlety, which will play a role in our story, is that $M$ must map normalizable $Q$-cohomology representatives to normalizable ones, otherwise it cannot be an isomorphism. 

\subsection{Deformations of BMN}
\label{sec:deform}

We want to find such a similarity transformation $M(\mu_{1},\mu_{0})$ for the BMN supercharges in eq. \eqref{eqn:supercharges}. Inspecting them, one finds that there is a single $M$ that works for all of them. Using the canonical commutation relations $[X_{nm}^{A},\Pi_{pq}^{B}] = i\delta^{AB}\delta_{nq}\delta_{mp}$, we obtain 
\begin{equation}
\label{conj3}
    M(\mu_{1},\mu_{0}) = e^{\frac{1}{6}(\mu_{1}-\mu_{0})\left(\frac{1}{2}\Tr(X_{a}X^{a})-\Tr(X_{i}X^{i})\right)}\,.
\end{equation}
Note that $H \propto \sum_{\alpha,I}\{Q^{\alpha I \dagger},Q_{\alpha I}\}$ so this changes the couplings in the Hamiltonian as well.

We need to make sure that the conjugation deformation preserves normalizability of the states. This requires estimating the fall-off of the wavefunctions at large $X_{A}$, which can be done using Agmon's estimate \cite{Agmon1982ExponentialDecay}. Given any solution to the $n$-dimensional Schr\"odinger's equation $\phi_{E}(x)$ with energy $E$, define the Agmon metric
\begin{equation}
    \rho_{E}(x,y) = \underset{\gamma}{\inf} \int_{0}^{1}dt\,|\dot{\gamma}|\,\sqrt{\max(0,V(\gamma(t))-E)}
\end{equation}
where $\gamma$ is a path from $\gamma(0)=x$ to $\gamma(1)=y$. Then the Agmon estimate says that the fall-off of the wavefunction at large $x$ (in the forbidden region) is, for any $\epsilon >0$,
\begin{equation}
    |\phi_{E}(x)| \leq \underset{\substack{y \in \mathbb{R}^{n} \\ V(y) \leq E}}{\sup} A_{\epsilon}e^{-(1-\epsilon)\rho_{E}(x,y)}\,.
\end{equation}
where $A_{\epsilon}$ is a constant depending only on $\epsilon$. For a heuristic derivation of this bound, see \cite{steinerberger2022agmonestimateschrodingeroperators}.

Now, let us apply this estimate to the BMN model. We care about the directions with the slowest fall-off since we are interested in normalizability. The weakest bound happens in the directions in $X_{A}$-space where the potential is the smallest at large $X_{A}$, which are the flat directions $[X_{A},X_{B}]=0$. In those directions, the potential is quadratic in $X_{A}$ so Agmon's estimate gives $|\phi_{E}(X)| \leq e^{-\mu c \Tr(X_{A}X^{A})}$ as $X_{A} \to \infty$ for some constant $c > 0$ independent of $\mu$ and $X$, where we have used large $x$ to drop $E$-dependent terms and the constant $A_{\epsilon}$. In this analysis, we have ignored the fermionic potential and the fermionic part of the Hilbert space because the fermionic terms, which are responsible for coupling together the different bosonic wavefunctions, are subleading at large $X_{A}$ so they can never affect our estimate.

Thus, we can now use an iterative argument to change $\mu$ \cite{WITTEN1982253}. Suppose we start at some $\mu_{0} > 0$ and we want to change it to $\mu_{1} > \mu_{0}$. Along the large $X_{i}$ directions, we find $M(\mu_{1},\mu_{0})^{-1} \sim e^{\frac{1}{6}(\mu_{1}-\mu_{0}) \Tr(X_{i}X^{i})}$ for $\mu_{1} > \mu_{0}$. When $\frac{1}{6}(\mu_{1}-\mu_{0}) > \mu_{0} c$, this will map our normalizable states at $\mu_{0}$ to non-normalizable states at $\mu_{1}$. The idea is to perform the change in coupling iteratively with each iteration preserving normalizability. There are, of course, infinitely many ways to do this but here is one way. Define the integer $m = \floor{\frac{\mu_{1}-\mu_{0}}{6\mu_{0}c}}$ and then use the conjugation deformation operator 
\begin{align}
\begin{split}
    C &= M\left(\mu_{1},\mu_{0}+2m\frac{6c\mu_{0}}{2}\right)
    \\  &\quad \times \prod_{k=1}^{2m}M\left(\mu_{0}+k\frac{6c\mu_{0}}{2},\mu_{0}+(k-1)\frac{6c\mu_{0}}{2}\right)\,.
\end{split}
\end{align}
Clearly, we can never get to the free theory $\mu_{1}=\infty$ by finitely many iterations, which had better be true since some BPS states were shown to lift perturbatively around the free point \cite{Dasgupta:2002ru, Kim:2002if}.

But what about if we want to go to smaller coupling $\mu_{1}<\mu_{0}$ and try to reach the BFSS point ($\mu_{1}=0$)? Now we need to check along large $X_{a}$ directions for normalizability, which gives $M(\mu_{1},\mu_{0})^{-1} \sim e^{\frac{1}{12}(\mu_{0}-\mu_{1}) \Tr(X_{a}X^{a})}$ for $\mu_{1} < \mu_{0}$. The problem with lowering $\mu$ is that the rate of decay of the wavefunction decreases after each iteration. For any $\mu_{1} > 0$, we can iterate with a step size of $|\delta \mu| < \mu_{1}$ and everything works. However, if we want to go all the way to $\mu_{1}=0$, we will have to take $|\delta \mu|$ smaller and smaller as we approach $\mu_{1}$, and hence we can never get there in finitely many iterations. This is consistent with the fact that the BPS spectrum of the BFSS model is different from the BMN one. Nevertheless, it must be the case that for some linear combination of the BPS vacua of BMN, the deformation $M(0,\mu_{0})$ acting on it is actually well-defined, since the BPS vacuum of BFSS exists.

\section{SHORT MULTIPLETS AND BPS States}
\label{sec:BPS}

\subsection{General structure}

In this section we review short representations of the BMN superalgebra and show that every short representation contains BPS states for one of the complex supercharges. Together with the conjugation argument presented in section \ref{sec:conjdef}, this implies that {\it  short multiplets in BMN cannot lift between any two finite non-vanishing values of $\mu$}. We emphasize that this is a non-perturbative statement and constrains the absolute number of short representations and thus goes beyond the signed-count constraint of the Witten index.

To understand the structure of representations, we think of the 8 complex $Q$'s of BMN which raise $H$ act as raising operators and the $Q^\dagger$'s as lowering operators. Each supermultiplet has a set of states of lowest energy, $E_0$, which are annihilated by all $Q^\dagger$, which we call {\it superprimaries}, and are 
characterized by an $SU(2)\times SU(4)$ representation $(j,[R_1,R_2,R_3])$. The rest of the multiplet is constructed by acting on these states with the 8 raising $Q$'s. Hence the entire supermulitplet is characterized by $(E_0;j,[R_1,R_2,R_3])$. For a generic long mulitplet we get $2^8\times {\rm dim}_{j,[R_1,R_2,R_3]}$ states.

Demanding unitarity (positivity of norms of super-descendants) introduces lower bounds for $E_0$ in a given charge sector. Multiplets which saturate this bound are ``short", or ``atypical". A classification of the short multiplets in BMN was presented in \cite{Dasgupta:2002ru} and we review it in appendix \ref{app:shortreps}. We have three general classes of shortening conditions. They are similar to the shortening conditions of local operators in the ${\cal N}=4$ SYM \cite{Dolan:2002zh, Kinney:2005ej, Cordova:2016emh} and we refer to them as of type $A_1,A_2,B$ \cite{Cordova:2016emh}.

In appendix \ref{app:shortreps} we also show that every possible short multiplet of the BMN superalgebra contains a BPS state for one of the complex supercharges of the theory. Some of these BPS states can never lift by representation theory, as shown in \cite{Dasgupta:2002ru}, while the others cannot lift between any finite, non-vanishing values of the coupling by our conjugation argument.

\subsection{Some comments on perturbative lifting}
\label{sec:pert}

Now let us discuss the lifting of states around the free point. The authors of \cite{Dasgupta:2002ru, Kim:2002if} identified an $A_{2}$-type short multiplet in the trivial vacuum $\mathcal{P}=(1,\ldots,1)$ whose states lift to second-order in perturbation theory around the free point, i.e fix $\mu$ and expand around $g=0$. The lowest energy state in this multiplet is an R-symmetry singlet and has energy $E_0={\mu\over 3}$, explicitly given by $|B\rangle=\mathrm{Tr}(A_a^{\dagger}A^{a\dagger})|0\rangle$ where $A_{a}^{\dagger}$ is the creation operator for $X_{a}$. This state is not BPS, but the supermultiplet contains 1/2-BPS states, such as $Q_{+1}Q_{+2}Q_{+3}Q_{+4}|B\rangle$ annihilated by all $Q_{-I}$ and $Q_{-I}^{\dagger}$. Therefore, our result would not apply to these states, as they would not survive at finite $\mu$.

On the other hand, it was noticed in \cite{Dasgupta:2002ru} that there are $A_{1}$-type short multiplets in $R$-symmetry representations $(2,[0,0,0])$ and $(\frac{3}{2},[1,0,0])$ in the irreducible vacuum $\mathcal{P}=(N)$ that could combine into a long multiplet and lift, but do not up to second-order in perturbation theory. These short multiplets thus contain candidate BPS states for which, if they do not lift to all orders or due to non-perturbative effects, so that they survive at small finite coupling, then our result would imply that they never lift even at strong coupling. 

\section{Chiral Ring}
\label{sec:chiral}
We now consider the cohomology of operators. We select one of the complex supercharges, say $Q=Q_{-,1}$, which satisfies $Q^2=0$. Its action on operators is defined by (graded) commutators. We define closed operators by
$$
[Q,{\cal O}]=0
$$
and identify
$$
{\cal O} \sim {\cal O} + [Q,X]\,.
$$
This defines cohomology classes of operators and their multiplication defines a ring. Notice that the ring is defined with respect to a particular supercharge $Q$.

In the BMN matrix model, examples of such chiral operators are in the multiplet of $S^{a_{1}\ldots a_{n}}\Tr(X_{a_{1}} \ldots X_{a_{n}})$ for $S$ a symmetric traceless tensor. Note that the chiral ring at the free point differs between the vacuum sectors because a different singlet condition is imposed on operators.

We can see that the ring is isomorphic for all finite, nonzero values of the coupling $\mu$, since the mapping ${\cal O}_i\rightarrow M^{-1} {\cal O}_i M$ maps operator cohomologies at different couplings and preserves the ring structure.

It is interesting to consider the matrix elements of $Q$-chiral operators on {\it states} of $Q$-cohomology,  namely the BPS states discussed in the previous subsection. 
Let $P_{\rm BPS}$ be the projector on the set of all BPS states. For each chiral operator ${\cal O}$ we consider \footnote{Notice that these have similarities to LMRS observables. In particular, we expect them to exhibit BPS chaos, at least for large enough R-charges and not too-supersymmetric states. This suggests that our conjugation deformation acts on the BPS subspace as a ``chaotic similarity transformation''.}
\be
\label{chirallmrs}
\widehat{{\cal O}} = P_{\rm BPS} {\cal O}P_{\rm PBS}\,.
\ee
While the operator $\Lambda ={1\over 6}\left(\frac{1}{2}\Tr(X_{a}X^{a})-\Tr(X_{i}X^{i})\right)$ generating the conjugation deformation is not chiral, we also define
\be
\widehat{\Lambda}= P_{\rm BPS} \Lambda P_{\rm PBS}\,.
\ee
Then a simple analysis, working in a choice of basis with vanishing Berry connection, gives
\be
{d\over d\mu}\widehat{\cal O} = \widehat{[\Lambda,{\cal O}]}-[\widehat{\Lambda},\widehat{\cal O}]\,.
\ee
Notice that in this equation $\widehat{\Lambda}$ is expected to be a non-trivial function of the coupling $\mu$. 

\section{Discussion}

\label{sec:disc}

One can ask if the conjugation deformation argument used here for the non-lifting of BPS states continues to hold in variants of the BMN model that preserve less supersymmetry, such as those studied in \cite{Motl:2003rw,Shimada:2008xy,Lee:2026xoo}. For the heterotic plane wave matrix model \cite{Motl:2003rw} obtained by quotienting the BMN model by a $\mathbb{Z}_{2}$ symmetry, our conjugation deformation operator in eq. \eqref{conj3} still works, and it would be nice to see how it gets modified in the other models.

An important open problem is to find the explicit BPS states---away from the free point---as function of $\mu$ and $N$. Recently, it was found that in $\mathcal{N}=4$ super-Yang Mills (SYM) theory, the BPS states fall into two different classes based on their behavior in $N$, known as the fortuity mechanism \cite{Chang:2022mjp, Chang:2024zqi}, and since generalized to many other theories \cite{Chang:2024lxt, Chang:2025wgo, Chang:2025wgo, Hughes:2025tdy, Belin:2025hsg, Behan:2025hbx}. The class of BPS states which exist only for $N \leq N_{\ast}$ for some $N_{\ast}$, dubbed fortuitous states, are conjectured to be dual to typical black hole microstates. In fact, the explicit construction of such states has thus far relied on the BMN truncation of $\mathcal{N}=4$ SYM \cite{Chang:2022mjp, Choi:2023znd, Choi:2023vdm, deMelloKoch:2024pcs}, which upon dimensional reduction reproduces the BMN matrix model classically and at one loop (although they differ at two-loop) \cite{Kim:2003rza}. Therefore, in the trivial vacuum, at least at one-loop, we expect to find the same structure for the fortuitous states. But for the other vacua, physical states are built out of $U(N_{1}) \times \ldots \times U(N_{k})$-invariant operators so it is a quiver gauge theory with the basic operators transforming in the bifundamental of $U(N_{i}) \times U(N_{j})$. We expect the fortuitous states in these vacua will have a similar structure to what one finds in ABJM \cite{Belin:2025hsg, Behan:2025hbx}. We hope to report on this in the near future.

It would also be interesting to understand if the BPS sector of the BMN model is chaotic, either in terms of eigenvalues of projected operators (LMRS notion of BPS chaos) \cite{Lin:2022zxd, Chen:2024oqv} or the Berry curvature \cite{Chen:2026vml}. One can argue, based on the fact that the model only has one effective coupling, that the Berry curvature trivially vanishes, but some of the variants have more couplings so the question could be interesting in those models. To our knowledge, it has not even been established whether the non-BPS part of the spectrum is quantum chaotic \cite{Asano:2015eha, Markovic:2022jta, Amore:2024ihm}, although general intuition for interacting many-body systems indicates that it should be. If so, how chaotic is it, namely does it saturate the Maldacena-Shenker-Stanford bound on the Lyapunov exponent \cite{Maldacena:2015waa} or have an $O(1)$ Thouless time?

\appendix
\section{A review of BMN short multiplets}
\label{app:shortreps}

Here we review the short multiplets of the BMN super-algebra \cite{Dasgupta:2002ru}, using similar conventions to \cite{Cordova:2016emh}. We also show that each short multiplet contains a BPS state which is a non-trivial cohomology class for one of the complex supercharges, which we take to be $Q_{-1}$.
We remind that
\be
\label{q1cohomology}
\{Q_{-1}^\dagger, Q_{-1}\}\equiv\Delta = 2H -{2\mu \over 3} j_3 -{\mu \over 6} (3R_1+2R_2+R_3)
\ee
and the condition $\Delta |\psi\rangle=0$ is equivalent to $Q|\psi\rangle= Q^{\dagger}|\psi\rangle=0$, i.e. the state $|\psi\rangle$ is a non-trivial cohomology element of $Q_{-1}$.

\subsection{Short reps with $j>0$, $A_1$ shortening}

Multiplets with $j>0$  must obey the bound\be
\label{bpsa1}
E_0 \geq {\mu\over 3}\left(j+1 +{3R_1+2R_2 +R_3 \over 4} \right)\,.
\ee
Consider a multiplet saturating this bound and let us denote by $|\phi_{+}^{j,j}\rangle$ a superprimary state in this multiplet, which is highest weight of $SU(4)$ and $SU(2)$. When the bound \eqref{bpsa1} is saturated we get null states already at the first level where we act with the raising operator $Q$ one time. The null states transform in the $(j-{1\over 2},[R_1+1,R_2,R_3])$ representation. The highest weight null state is
\be
\label{nulla1}
Q_{-1}|\phi^{j,j}_+\rangle -{1\over \sqrt{2j}} Q_{+1}|\phi^{j,j-1}\rangle =0\,.
\ee
Notice that these multiplets are sitting at the bottom of the continuum and can in principle recombine into long multiplets.

These multiplets contain BPS states in $Q_{-1}$ cohomology. For example, consider the state 
\be
|\psi\rangle = Q_{+1}|\phi_+^{jj}\rangle\,.
\ee
The state $|\phi_+^{jj}\rangle$ has charges $E_0,j,[R_1,R_2,R_3]$. The state $|\psi\rangle$ has $E_0+{\mu\over 12},j+{1\over 2},[R_1+1,R_2,R_3]$. If \eqref{bpsa1} is saturated we find from \eqref{q1cohomology} that $\Delta_{|\psi\rangle}=0$. It is indeed easy to check that $Q_{-1}|\psi\rangle=Q_{-1}^\dagger|\psi\rangle=0$, hence we have a non-trivial BPS state.

\subsection{Short reps with $j=0$, $A_2$ shortening}

When $j=0$, a unitarity bound coming from 2nd descendants reads
\be
\label{bpsa2}
E_0\geq {\mu \over 3} \left(1+{3R_1+2R_2+R_3\over4}\right)\,.
\ee
Let us denote by $|\phi_+\rangle$ the superprimary which is highest weight of $SU(4)$. If \eqref{bpsa2} is saturated we find the following null state
\be
Q_{-1} Q_{+1} |\phi_+\rangle=0
\ee
which has $j=0,[R_1+2,R_2,R_3]$. Again, these multiplets sit at the bottom of a continuum of long multiplets and can continuously lift. 

This multiplet also contains $Q_{-1}$ BPS states. Consider the state
\be
|\psi\rangle= Q_{+1}|\phi_+\rangle
\ee
This state has charges $E_0+{\mu\over 12}, j_3={1\over 2}, [R_1+1,R_2,R_3]$. If \eqref{bpsa2} is saturated and evaluating \eqref{q1cohomology} we again find $\Delta_{|\psi\rangle}=0$, so we have $Q_{-1}|\psi\rangle = Q_{-1}^\dagger|\psi\rangle=0$.

\subsection{Short reps with $j=0$, $B$-shortening}

When $j=0$ there is yet another class of unitary, short representations which saturate
\be
\label{bpsb}
E_0= {\mu \over 3}\left({3R_1+2R_2 +R_3 \over 4}\right)\,.
\ee
Notice that these are separated by a gap of width ${\mu \over 3}$ from the continuum. The ones with $R_1=0$ cannot continuously recombine into long multiplets and are thus absolutely protected --- these representations were called doubly atypical in \cite{Dasgupta:2002ru}.

Let us denote by $|\phi_+\rangle$ the superprimary $SU(2)\times SU(4)$ highest weight state. We have the null-ness condition
\be
Q_{\pm 1} |\phi_+\rangle=0\,.
\ee
Since $|\phi_+\rangle$ has the lowest possible energy in the multiplet it is also annihilated by $Q_{\pm 1}^\dagger$, hence is $Q_{-1}$ BPS.

\section{Failure at the free point}

We explained in Sec. \ref{sec:deform} why one cannot reach the free point using conjugation deformations by starting at finite $\mu>0$. Instead, one could try to start at the free point and deform away \emph{perturbatively} using conjugation deformations. This can still be useful even when $M(\mu,0)=e^{\mu\Lambda}$ is not a ``good'' operator, since the generator $\Lambda$ of the deformation may still be. It was shown in \cite{WITTEN1982253} that when such a $\Lambda$ exists and $[\Lambda,[\Lambda,Q]]=0$ (and the same for $Q^{\dagger}$), then BPS states cannot lift to any order in perturbation theory. But, as discussed in Sec. \ref{sec:pert}, there are BPS states that lift after perturbing away from the point, so there must not be a permissible $\Lambda$. The goal of this Appendix is to check this explicitly.

Perturbing around the free point means fixing $\mu$ and expanding around $g=0$. The variation of the supercharge is
\begin{align}
\begin{split}
\delta_{g}Q_{\alpha I} &= \Tr \bigg(\frac{1}{2} [X_i, X_j] \epsilon^{ijk} \sigma^k_\alpha {}^\beta 
\psi_{I \beta} -\frac{i}{2} [X_a, X_b] ({\sf g}^{ab})_I^J \psi_{J \alpha} 
\\  &\qquad +i [X_i, X_a] \sigma^i_\alpha {}^\lambda {\sf g}^a_{IJ} \epsilon_{\lambda \beta} \psi^{\dagger J 
\beta} \bigg)\,,
\end{split}
\end{align}
and we want to find $\widetilde{\Lambda}$ such that
\begin{equation}
\label{eqn:Qpert}
    [Q_{\alpha I},\widetilde{\Lambda}] = \delta_{g}Q_{\alpha I}\,.
\end{equation}
By inspection, we find that the only possibility is that it takes the form $\widetilde{\Lambda} = R_{ABC}\Tr(X^{A}X^{B}X^{C})$. However, this does not work because the $\sf g$ and $\sigma$ matrices do not match on the two sides of eq. \eqref{eqn:Qpert}. The only option is thus to promote $\widetilde{\Lambda}$ to a matrix acting on the $SU(2) \times SU(4)$ indices, i.e. $R_{ABC} \to (R_{ABC})_{\beta J}^{\alpha I}$, then find such a $\widetilde{\Lambda}$ that solves $[Q_{\alpha I},(\widetilde{\Lambda})_{\beta J}^{\alpha I}] = \delta_{g}Q_{\beta J}$, and try to diagonalize this equation. We have checked that a solution can never exist, and hence perturbative changes of the coupling fail, as expected.

\bibliography{Refs}

\begin{thebibliography}{45}%
\makeatletter
\providecommand \@ifxundefined [1]{%
 \@ifx{#1\undefined}
}%
\providecommand \@ifnum [1]{%
 \ifnum #1\expandafter \@firstoftwo
 \else \expandafter \@secondoftwo
 \fi
}%
\providecommand \@ifx [1]{%
 \ifx #1\expandafter \@firstoftwo
 \else \expandafter \@secondoftwo
 \fi
}%
\providecommand \natexlab [1]{#1}%
\providecommand \enquote  [1]{``#1''}%
\providecommand \bibnamefont  [1]{#1}%
\providecommand \bibfnamefont [1]{#1}%
\providecommand \citenamefont [1]{#1}%
\providecommand \href@noop [0]{\@secondoftwo}%
\providecommand \href [0]{\begingroup \@sanitize@url \@href}%
\providecommand \@href[1]{\@@startlink{#1}\@@href}%
\providecommand \@@href[1]{\endgroup#1\@@endlink}%
\providecommand \@sanitize@url [0]{\catcode `\\12\catcode `\$12\catcode `\&12\catcode `\#12\catcode `\^12\catcode `\_12\catcode `\%12\relax}%
\providecommand \@@startlink[1]{}%
\providecommand \@@endlink[0]{}%
\providecommand \url  [0]{\begingroup\@sanitize@url \@url }%
\providecommand \@url [1]{\endgroup\@href {#1}{\urlprefix }}%
\providecommand \urlprefix  [0]{URL }%
\providecommand \Eprint [0]{\href }%
\providecommand \doibase [0]{https://doi.org/}%
\providecommand \selectlanguage [0]{\@gobble}%
\providecommand \bibinfo  [0]{\@secondoftwo}%
\providecommand \bibfield  [0]{\@secondoftwo}%
\providecommand \translation [1]{[#1]}%
\providecommand \BibitemOpen [0]{}%
\providecommand \bibitemStop [0]{}%
\providecommand \bibitemNoStop [0]{.\EOS\space}%
\providecommand \EOS [0]{\spacefactor3000\relax}%
\providecommand \BibitemShut  [1]{\csname bibitem#1\endcsname}%
\let\auto@bib@innerbib\@empty
\bibitem [{\citenamefont {Berenstein}\ \emph {et~al.}(2002)\citenamefont {Berenstein}, \citenamefont {Maldacena},\ and\ \citenamefont {Nastase}}]{Berenstein:2002jq}%
  \BibitemOpen
  \bibfield  {author} {\bibinfo {author} {\bibfnamefont {D.~E.}\ \bibnamefont {Berenstein}}, \bibinfo {author} {\bibfnamefont {J.~M.}\ \bibnamefont {Maldacena}},\ and\ \bibinfo {author} {\bibfnamefont {H.~S.}\ \bibnamefont {Nastase}},\ }\bibfield  {title} {\bibinfo {title} {{Strings in flat space and pp waves from N=4 superYang-Mills}},\ }\href {https://doi.org/10.1088/1126-6708/2002/04/013} {\bibfield  {journal} {\bibinfo  {journal} {JHEP}\ }\textbf {\bibinfo {volume} {04}},\ \bibinfo {pages} {013}},\ \Eprint {https://arxiv.org/abs/hep-th/0202021} {arXiv:hep-th/0202021} \BibitemShut {NoStop}%
\bibitem [{\citenamefont {Costa}\ \emph {et~al.}(2015)\citenamefont {Costa}, \citenamefont {Greenspan}, \citenamefont {Penedones},\ and\ \citenamefont {Santos}}]{Costa:2014wya}%
  \BibitemOpen
  \bibfield  {author} {\bibinfo {author} {\bibfnamefont {M.~S.}\ \bibnamefont {Costa}}, \bibinfo {author} {\bibfnamefont {L.}~\bibnamefont {Greenspan}}, \bibinfo {author} {\bibfnamefont {J.}~\bibnamefont {Penedones}},\ and\ \bibinfo {author} {\bibfnamefont {J.}~\bibnamefont {Santos}},\ }\bibfield  {title} {\bibinfo {title} {{Thermodynamics of the BMN matrix model at strong coupling}},\ }\href {https://doi.org/10.1007/JHEP03(2015)069} {\bibfield  {journal} {\bibinfo  {journal} {JHEP}\ }\textbf {\bibinfo {volume} {03}},\ \bibinfo {pages} {069}},\ \Eprint {https://arxiv.org/abs/1411.5541} {arXiv:1411.5541 [hep-th]} \BibitemShut {NoStop}%
\bibitem [{\citenamefont {Chang}(2025)}]{Chang:2024lkw}%
  \BibitemOpen
  \bibfield  {author} {\bibinfo {author} {\bibfnamefont {C.-M.}\ \bibnamefont {Chang}},\ }\bibfield  {title} {\bibinfo {title} {{Witten index of BMN matrix quantum mechanics}},\ }\href {https://doi.org/10.21468/SciPostPhys.19.6.147} {\bibfield  {journal} {\bibinfo  {journal} {SciPost Phys.}\ }\textbf {\bibinfo {volume} {19}},\ \bibinfo {pages} {147} (\bibinfo {year} {2025})},\ \Eprint {https://arxiv.org/abs/2404.18442} {arXiv:2404.18442 [hep-th]} \BibitemShut {NoStop}%
\bibitem [{\citenamefont {Zigdon}(2026)}]{Zigdon:2025icw}%
  \BibitemOpen
  \bibfield  {author} {\bibinfo {author} {\bibfnamefont {Y.}~\bibnamefont {Zigdon}},\ }\bibfield  {title} {\bibinfo {title} {{Charge constraint in the Berenstein-Maldacena-Nastase model}},\ }\href {https://doi.org/10.1103/cl75-thnk} {\bibfield  {journal} {\bibinfo  {journal} {Phys. Rev. D}\ }\textbf {\bibinfo {volume} {113}},\ \bibinfo {pages} {026002} (\bibinfo {year} {2026})},\ \Eprint {https://arxiv.org/abs/2506.19924} {arXiv:2506.19924 [hep-th]} \BibitemShut {NoStop}%
\bibitem [{\citenamefont {Chang}\ \emph {et~al.}(2026)\citenamefont {Chang}, \citenamefont {Duary},\ and\ \citenamefont {Liu}}]{Chang:2026mui}%
  \BibitemOpen
  \bibfield  {author} {\bibinfo {author} {\bibfnamefont {C.-M.}\ \bibnamefont {Chang}}, \bibinfo {author} {\bibfnamefont {S.}~\bibnamefont {Duary}},\ and\ \bibinfo {author} {\bibfnamefont {K.}~\bibnamefont {Liu}},\ }\bibfield  {title} {\bibinfo {title} {{Finite-$N$ BMN index across all vacuum sectors}},\ }\href@noop {} {\  (\bibinfo {year} {2026})},\ \Eprint {https://arxiv.org/abs/2605.25560} {arXiv:2605.25560 [hep-th]} \BibitemShut {NoStop}%
\bibitem [{Note1()}]{Note1}%
  \BibitemOpen
  \bibinfo {note} {This Hamiltonian is related to the standard one in eq. (1) of \cite {Dasgupta:2002ru} by $g=R^{\protect \frac {3}{2}}$ and the redefinition $\Pi ^{\protect \mathrm {us}} = g^{\protect \frac {1}{3}}\Pi ^{\protect \mathrm {them}}$ and $X^{\protect \mathrm {us}} = g^{-\protect \frac {1}{3}}X^{\protect \mathrm {them}}$. Our choice gives a canonically normalized kinetic term. We use the same conventions as them for spinors, gamma matrices, etc.}\BibitemShut {Stop}%
\bibitem [{Note2()}]{Note2}%
  \BibitemOpen
  \bibinfo {note} {There is a trivial $U(1)$ sector of the model coming from the trace mode, which is always free, and has an extra 16 supercharges associated to it. The non-trivial, interacting $SU(N)$ sector consists of traceless, Hermitian matrices. We will restrict attention to the non-trivial part of the theory.}\BibitemShut {Stop}%
\bibitem [{\citenamefont {Banks}\ \emph {et~al.}(1997)\citenamefont {Banks}, \citenamefont {Fischler}, \citenamefont {Shenker},\ and\ \citenamefont {Susskind}}]{Banks:1996vh}%
  \BibitemOpen
  \bibfield  {author} {\bibinfo {author} {\bibfnamefont {T.}~\bibnamefont {Banks}}, \bibinfo {author} {\bibfnamefont {W.}~\bibnamefont {Fischler}}, \bibinfo {author} {\bibfnamefont {S.~H.}\ \bibnamefont {Shenker}},\ and\ \bibinfo {author} {\bibfnamefont {L.}~\bibnamefont {Susskind}},\ }\bibfield  {title} {\bibinfo {title} {{M theory as a matrix model: A conjecture}},\ }\href {https://doi.org/10.1103/PhysRevD.55.5112} {\bibfield  {journal} {\bibinfo  {journal} {Phys. Rev. D}\ }\textbf {\bibinfo {volume} {55}},\ \bibinfo {pages} {5112} (\bibinfo {year} {1997})},\ \Eprint {https://arxiv.org/abs/hep-th/9610043} {arXiv:hep-th/9610043} \BibitemShut {NoStop}%
\bibitem [{\citenamefont {{De Wit}}\ \emph {et~al.}(1989)\citenamefont {{De Wit}}, \citenamefont {L{\"u}scher},\ and\ \citenamefont {Nicolai}}]{DEWIT1989135}%
  \BibitemOpen
  \bibfield  {author} {\bibinfo {author} {\bibfnamefont {B.}~\bibnamefont {{De Wit}}}, \bibinfo {author} {\bibfnamefont {M.}~\bibnamefont {L{\"u}scher}},\ and\ \bibinfo {author} {\bibfnamefont {H.}~\bibnamefont {Nicolai}},\ }\bibfield  {title} {\bibinfo {title} {The supermembrane is unstable},\ }\href {https://doi.org/https://doi.org/10.1016/0550-3213(89)90214-9} {\bibfield  {journal} {\bibinfo  {journal} {Nuclear Physics B}\ }\textbf {\bibinfo {volume} {320}},\ \bibinfo {pages} {135} (\bibinfo {year} {1989})}\BibitemShut {NoStop}%
\bibitem [{\citenamefont {Sethi}\ and\ \citenamefont {Stern}(1998)}]{Sethi:1997pa}%
  \BibitemOpen
  \bibfield  {author} {\bibinfo {author} {\bibfnamefont {S.}~\bibnamefont {Sethi}}\ and\ \bibinfo {author} {\bibfnamefont {M.}~\bibnamefont {Stern}},\ }\bibfield  {title} {\bibinfo {title} {{D-brane bound states redux}},\ }\href {https://doi.org/10.1007/s002200050374} {\bibfield  {journal} {\bibinfo  {journal} {Commun. Math. Phys.}\ }\textbf {\bibinfo {volume} {194}},\ \bibinfo {pages} {675} (\bibinfo {year} {1998})},\ \Eprint {https://arxiv.org/abs/hep-th/9705046} {arXiv:hep-th/9705046} \BibitemShut {NoStop}%
\bibitem [{\citenamefont {Porrati}\ and\ \citenamefont {Rozenberg}(1998)}]{Porrati:1997ej}%
  \BibitemOpen
  \bibfield  {author} {\bibinfo {author} {\bibfnamefont {M.}~\bibnamefont {Porrati}}\ and\ \bibinfo {author} {\bibfnamefont {A.}~\bibnamefont {Rozenberg}},\ }\bibfield  {title} {\bibinfo {title} {{Bound states at threshold in supersymmetric quantum mechanics}},\ }\href {https://doi.org/10.1016/S0550-3213(97)00804-3} {\bibfield  {journal} {\bibinfo  {journal} {Nucl. Phys. B}\ }\textbf {\bibinfo {volume} {515}},\ \bibinfo {pages} {184} (\bibinfo {year} {1998})},\ \Eprint {https://arxiv.org/abs/hep-th/9708119} {arXiv:hep-th/9708119} \BibitemShut {NoStop}%
\bibitem [{\citenamefont {Moore}\ \emph {et~al.}(2000)\citenamefont {Moore}, \citenamefont {Nekrasov},\ and\ \citenamefont {Shatashvili}}]{Moore:1998et}%
  \BibitemOpen
  \bibfield  {author} {\bibinfo {author} {\bibfnamefont {G.~W.}\ \bibnamefont {Moore}}, \bibinfo {author} {\bibfnamefont {N.}~\bibnamefont {Nekrasov}},\ and\ \bibinfo {author} {\bibfnamefont {S.}~\bibnamefont {Shatashvili}},\ }\bibfield  {title} {\bibinfo {title} {{D-particle bound states and generalized instantons}},\ }\href {https://doi.org/10.1007/s002200050016} {\bibfield  {journal} {\bibinfo  {journal} {Commun. Math. Phys.}\ }\textbf {\bibinfo {volume} {209}},\ \bibinfo {pages} {77} (\bibinfo {year} {2000})},\ \Eprint {https://arxiv.org/abs/hep-th/9803265} {arXiv:hep-th/9803265} \BibitemShut {NoStop}%
\bibitem [{\citenamefont {Dasgupta}\ \emph {et~al.}(2002{\natexlab{a}})\citenamefont {Dasgupta}, \citenamefont {Sheikh-Jabbari},\ and\ \citenamefont {Van~Raamsdonk}}]{Dasgupta:2002hx}%
  \BibitemOpen
  \bibfield  {author} {\bibinfo {author} {\bibfnamefont {K.}~\bibnamefont {Dasgupta}}, \bibinfo {author} {\bibfnamefont {M.~M.}\ \bibnamefont {Sheikh-Jabbari}},\ and\ \bibinfo {author} {\bibfnamefont {M.}~\bibnamefont {Van~Raamsdonk}},\ }\bibfield  {title} {\bibinfo {title} {{Matrix perturbation theory for M theory on a PP wave}},\ }\href {https://doi.org/10.1088/1126-6708/2002/05/056} {\bibfield  {journal} {\bibinfo  {journal} {JHEP}\ }\textbf {\bibinfo {volume} {05}},\ \bibinfo {pages} {056}},\ \Eprint {https://arxiv.org/abs/hep-th/0205185} {arXiv:hep-th/0205185} \BibitemShut {NoStop}%
\bibitem [{\citenamefont {Dasgupta}\ \emph {et~al.}(2002{\natexlab{b}})\citenamefont {Dasgupta}, \citenamefont {Sheikh-Jabbari},\ and\ \citenamefont {Van~Raamsdonk}}]{Dasgupta:2002ru}%
  \BibitemOpen
  \bibfield  {author} {\bibinfo {author} {\bibfnamefont {K.}~\bibnamefont {Dasgupta}}, \bibinfo {author} {\bibfnamefont {M.~M.}\ \bibnamefont {Sheikh-Jabbari}},\ and\ \bibinfo {author} {\bibfnamefont {M.}~\bibnamefont {Van~Raamsdonk}},\ }\bibfield  {title} {\bibinfo {title} {{Protected multiplets of M theory on a plane wave}},\ }\href {https://doi.org/10.1088/1126-6708/2002/09/021} {\bibfield  {journal} {\bibinfo  {journal} {JHEP}\ }\textbf {\bibinfo {volume} {09}},\ \bibinfo {pages} {021}},\ \Eprint {https://arxiv.org/abs/hep-th/0207050} {arXiv:hep-th/0207050} \BibitemShut {NoStop}%
\bibitem [{\citenamefont {Kim}\ and\ \citenamefont {Park}(2002)}]{Kim:2002zg}%
  \BibitemOpen
  \bibfield  {author} {\bibinfo {author} {\bibfnamefont {N.}~\bibnamefont {Kim}}\ and\ \bibinfo {author} {\bibfnamefont {J.-H.}\ \bibnamefont {Park}},\ }\bibfield  {title} {\bibinfo {title} {{Superalgebra for M theory on a pp wave}},\ }\href {https://doi.org/10.1103/PhysRevD.66.106007} {\bibfield  {journal} {\bibinfo  {journal} {Phys. Rev. D}\ }\textbf {\bibinfo {volume} {66}},\ \bibinfo {pages} {106007} (\bibinfo {year} {2002})},\ \Eprint {https://arxiv.org/abs/hep-th/0207061} {arXiv:hep-th/0207061} \BibitemShut {NoStop}%
\bibitem [{Note3()}]{Note3}%
  \BibitemOpen
  \bibinfo {note} {Note that in the literature sometimes the 16 supercharges from the trivial $U(1)$ sector are also included in the counting, so the designation of BPS is $1/2$ of ours, e.g. the vacuum would be 1/2-BPS in those conventions while it is fully BPS in ours.}\BibitemShut {Stop}%
\bibitem [{\citenamefont {Witten}(1982)}]{WITTEN1982253}%
  \BibitemOpen
  \bibfield  {author} {\bibinfo {author} {\bibfnamefont {E.}~\bibnamefont {Witten}},\ }\bibfield  {title} {\bibinfo {title} {Constraints on supersymmetry breaking},\ }\href {https://doi.org/https://doi.org/10.1016/0550-3213(82)90071-2} {\bibfield  {journal} {\bibinfo  {journal} {Nuclear Physics B}\ }\textbf {\bibinfo {volume} {202}},\ \bibinfo {pages} {253} (\bibinfo {year} {1982})}\BibitemShut {NoStop}%
\bibitem [{\citenamefont {Agmon}(1982)}]{Agmon1982ExponentialDecay}%
  \BibitemOpen
  \bibfield  {author} {\bibinfo {author} {\bibfnamefont {S.}~\bibnamefont {Agmon}},\ }\href@noop {} {\emph {\bibinfo {title} {Lectures on Exponential Decay of Solutions of Second-Order Elliptic Equations: Bounds on Eigenfunctions of {N}-Body Schrödinger Operators}}},\ \bibinfo {series} {Mathematical Notes}, Vol.~\bibinfo {volume} {29}\ (\bibinfo  {publisher} {Princeton University Press},\ \bibinfo {address} {Princeton, NJ},\ \bibinfo {year} {1982})\BibitemShut {NoStop}%
\bibitem [{\citenamefont {Steinerberger}(2022)}]{steinerberger2022agmonestimateschrodingeroperators}%
  \BibitemOpen
  \bibfield  {author} {\bibinfo {author} {\bibfnamefont {S.}~\bibnamefont {Steinerberger}},\ }\href {https://arxiv.org/abs/2206.09521} {\bibinfo {title} {{An Agmon estimate for Schr\"odinger operators on Graphs}}} (\bibinfo {year} {2022}),\ \Eprint {https://arxiv.org/abs/2206.09521} {arXiv:2206.09521 [math.SP]} \BibitemShut {NoStop}%
\bibitem [{\citenamefont {Kim}\ and\ \citenamefont {Plefka}(2002)}]{Kim:2002if}%
  \BibitemOpen
  \bibfield  {author} {\bibinfo {author} {\bibfnamefont {N.}~\bibnamefont {Kim}}\ and\ \bibinfo {author} {\bibfnamefont {J.}~\bibnamefont {Plefka}},\ }\bibfield  {title} {\bibinfo {title} {{On the spectrum of PP wave matrix theory}},\ }\href {https://doi.org/10.1016/S0550-3213(02)00738-1} {\bibfield  {journal} {\bibinfo  {journal} {Nucl. Phys. B}\ }\textbf {\bibinfo {volume} {643}},\ \bibinfo {pages} {31} (\bibinfo {year} {2002})},\ \Eprint {https://arxiv.org/abs/hep-th/0207034} {arXiv:hep-th/0207034} \BibitemShut {NoStop}%
\bibitem [{\citenamefont {Dolan}\ and\ \citenamefont {Osborn}(2003)}]{Dolan:2002zh}%
  \BibitemOpen
  \bibfield  {author} {\bibinfo {author} {\bibfnamefont {F.~A.}\ \bibnamefont {Dolan}}\ and\ \bibinfo {author} {\bibfnamefont {H.}~\bibnamefont {Osborn}},\ }\bibfield  {title} {\bibinfo {title} {{On short and semi-short representations for four-dimensional superconformal symmetry}},\ }\href {https://doi.org/10.1016/S0003-4916(03)00074-5} {\bibfield  {journal} {\bibinfo  {journal} {Annals Phys.}\ }\textbf {\bibinfo {volume} {307}},\ \bibinfo {pages} {41} (\bibinfo {year} {2003})},\ \Eprint {https://arxiv.org/abs/hep-th/0209056} {arXiv:hep-th/0209056} \BibitemShut {NoStop}%
\bibitem [{\citenamefont {Kinney}\ \emph {et~al.}(2007)\citenamefont {Kinney}, \citenamefont {Maldacena}, \citenamefont {Minwalla},\ and\ \citenamefont {Raju}}]{Kinney:2005ej}%
  \BibitemOpen
  \bibfield  {author} {\bibinfo {author} {\bibfnamefont {J.}~\bibnamefont {Kinney}}, \bibinfo {author} {\bibfnamefont {J.~M.}\ \bibnamefont {Maldacena}}, \bibinfo {author} {\bibfnamefont {S.}~\bibnamefont {Minwalla}},\ and\ \bibinfo {author} {\bibfnamefont {S.}~\bibnamefont {Raju}},\ }\bibfield  {title} {\bibinfo {title} {{An Index for 4 dimensional super conformal theories}},\ }\href {https://doi.org/10.1007/s00220-007-0258-7} {\bibfield  {journal} {\bibinfo  {journal} {Commun. Math. Phys.}\ }\textbf {\bibinfo {volume} {275}},\ \bibinfo {pages} {209} (\bibinfo {year} {2007})},\ \Eprint {https://arxiv.org/abs/hep-th/0510251} {arXiv:hep-th/0510251} \BibitemShut {NoStop}%
\bibitem [{\citenamefont {Cordova}\ \emph {et~al.}(2019)\citenamefont {Cordova}, \citenamefont {Dumitrescu},\ and\ \citenamefont {Intriligator}}]{Cordova:2016emh}%
  \BibitemOpen
  \bibfield  {author} {\bibinfo {author} {\bibfnamefont {C.}~\bibnamefont {Cordova}}, \bibinfo {author} {\bibfnamefont {T.~T.}\ \bibnamefont {Dumitrescu}},\ and\ \bibinfo {author} {\bibfnamefont {K.}~\bibnamefont {Intriligator}},\ }\bibfield  {title} {\bibinfo {title} {{Multiplets of Superconformal Symmetry in Diverse Dimensions}},\ }\href {https://doi.org/10.1007/JHEP03(2019)163} {\bibfield  {journal} {\bibinfo  {journal} {JHEP}\ }\textbf {\bibinfo {volume} {03}},\ \bibinfo {pages} {163}},\ \Eprint {https://arxiv.org/abs/1612.00809} {arXiv:1612.00809 [hep-th]} \BibitemShut {NoStop}%
\bibitem [{Note4()}]{Note4}%
  \BibitemOpen
  \bibinfo {note} {Notice that these have similarities to LMRS observables. In particular, we expect them to exhibit BPS chaos, at least for large enough R-charges and not too-supersymmetric states. This suggests that our conjugation deformation acts on the BPS subspace as a ``chaotic similarity transformation''.}\BibitemShut {Stop}%
\bibitem [{\citenamefont {Motl}\ \emph {et~al.}(2003)\citenamefont {Motl}, \citenamefont {Neitzke},\ and\ \citenamefont {Sheikh-Jabbari}}]{Motl:2003rw}%
  \BibitemOpen
  \bibfield  {author} {\bibinfo {author} {\bibfnamefont {L.}~\bibnamefont {Motl}}, \bibinfo {author} {\bibfnamefont {A.}~\bibnamefont {Neitzke}},\ and\ \bibinfo {author} {\bibfnamefont {M.~M.}\ \bibnamefont {Sheikh-Jabbari}},\ }\bibfield  {title} {\bibinfo {title} {{Heterotic plane wave matrix models and giant gluons}},\ }\href {https://doi.org/10.1088/1126-6708/2003/06/058} {\bibfield  {journal} {\bibinfo  {journal} {JHEP}\ }\textbf {\bibinfo {volume} {06}},\ \bibinfo {pages} {058}},\ \Eprint {https://arxiv.org/abs/hep-th/0306051} {arXiv:hep-th/0306051} \BibitemShut {NoStop}%
\bibitem [{\citenamefont {Shimada}(2009)}]{Shimada:2008xy}%
  \BibitemOpen
  \bibfield  {author} {\bibinfo {author} {\bibfnamefont {H.}~\bibnamefont {Shimada}},\ }\bibfield  {title} {\bibinfo {title} {{beta-deformation for matrix model of M-theory}},\ }\href {https://doi.org/10.1016/j.nuclphysb.2008.08.018} {\bibfield  {journal} {\bibinfo  {journal} {Nucl. Phys. B}\ }\textbf {\bibinfo {volume} {813}},\ \bibinfo {pages} {283} (\bibinfo {year} {2009})},\ \Eprint {https://arxiv.org/abs/0804.3236} {arXiv:0804.3236 [hep-th]} \BibitemShut {NoStop}%
\bibitem [{\citenamefont {Lee}(2026)}]{Lee:2026xoo}%
  \BibitemOpen
  \bibfield  {author} {\bibinfo {author} {\bibfnamefont {E.}~\bibnamefont {Lee}},\ }\bibfield  {title} {\bibinfo {title} {{BMN-like Matrix Models}},\ }\href@noop {} {\  (\bibinfo {year} {2026})},\ \Eprint {https://arxiv.org/abs/2602.22163} {arXiv:2602.22163 [hep-th]} \BibitemShut {NoStop}%
\bibitem [{\citenamefont {Chang}\ and\ \citenamefont {Lin}(2023)}]{Chang:2022mjp}%
  \BibitemOpen
  \bibfield  {author} {\bibinfo {author} {\bibfnamefont {C.-M.}\ \bibnamefont {Chang}}\ and\ \bibinfo {author} {\bibfnamefont {Y.-H.}\ \bibnamefont {Lin}},\ }\bibfield  {title} {\bibinfo {title} {{Words to describe a black hole}},\ }\href {https://doi.org/10.1007/JHEP02(2023)109} {\bibfield  {journal} {\bibinfo  {journal} {JHEP}\ }\textbf {\bibinfo {volume} {02}},\ \bibinfo {pages} {109}},\ \Eprint {https://arxiv.org/abs/2209.06728} {arXiv:2209.06728 [hep-th]} \BibitemShut {NoStop}%
\bibitem [{\citenamefont {Chang}\ and\ \citenamefont {Lin}(2024)}]{Chang:2024zqi}%
  \BibitemOpen
  \bibfield  {author} {\bibinfo {author} {\bibfnamefont {C.-M.}\ \bibnamefont {Chang}}\ and\ \bibinfo {author} {\bibfnamefont {Y.-H.}\ \bibnamefont {Lin}},\ }\bibfield  {title} {\bibinfo {title} {{Holographic covering and the fortuity of black holes}},\ }\href@noop {} {\  (\bibinfo {year} {2024})},\ \Eprint {https://arxiv.org/abs/2402.10129} {arXiv:2402.10129 [hep-th]} \BibitemShut {NoStop}%
\bibitem [{\citenamefont {Chang}\ \emph {et~al.}(2025)\citenamefont {Chang}, \citenamefont {Chen}, \citenamefont {Sia},\ and\ \citenamefont {Yang}}]{Chang:2024lxt}%
  \BibitemOpen
  \bibfield  {author} {\bibinfo {author} {\bibfnamefont {C.-M.}\ \bibnamefont {Chang}}, \bibinfo {author} {\bibfnamefont {Y.}~\bibnamefont {Chen}}, \bibinfo {author} {\bibfnamefont {B.~S.}\ \bibnamefont {Sia}},\ and\ \bibinfo {author} {\bibfnamefont {Z.}~\bibnamefont {Yang}},\ }\bibfield  {title} {\bibinfo {title} {{Fortuity in SYK models}},\ }\href {https://doi.org/10.1007/JHEP08(2025)003} {\bibfield  {journal} {\bibinfo  {journal} {JHEP}\ }\textbf {\bibinfo {volume} {08}},\ \bibinfo {pages} {003}},\ \Eprint {https://arxiv.org/abs/2412.06902} {arXiv:2412.06902 [hep-th]} \BibitemShut {NoStop}%
\bibitem [{\citenamefont {Chang}\ and\ \citenamefont {Zhang}(2025)}]{Chang:2025wgo}%
  \BibitemOpen
  \bibfield  {author} {\bibinfo {author} {\bibfnamefont {C.-M.}\ \bibnamefont {Chang}}\ and\ \bibinfo {author} {\bibfnamefont {H.}~\bibnamefont {Zhang}},\ }\bibfield  {title} {\bibinfo {title} {{Fortuity and R-charge concentration in the D1-D5 CFT}},\ }\href@noop {} {\  (\bibinfo {year} {2025})},\ \Eprint {https://arxiv.org/abs/2511.23294} {arXiv:2511.23294 [hep-th]} \BibitemShut {NoStop}%
\bibitem [{\citenamefont {Hughes}\ and\ \citenamefont {Shigemori}(2026)}]{Hughes:2025tdy}%
  \BibitemOpen
  \bibfield  {author} {\bibinfo {author} {\bibfnamefont {M.~R.~R.}\ \bibnamefont {Hughes}}\ and\ \bibinfo {author} {\bibfnamefont {M.}~\bibnamefont {Shigemori}},\ }\bibfield  {title} {\bibinfo {title} {{Fortuity and supergravity}},\ }\href {https://doi.org/10.1007/JHEP03(2026)130} {\bibfield  {journal} {\bibinfo  {journal} {JHEP}\ }\textbf {\bibinfo {volume} {03}},\ \bibinfo {pages} {130}},\ \Eprint {https://arxiv.org/abs/2505.14888} {arXiv:2505.14888 [hep-th]} \BibitemShut {NoStop}%
\bibitem [{\citenamefont {Belin}\ \emph {et~al.}(2025)\citenamefont {Belin}, \citenamefont {Singh}, \citenamefont {Vadala},\ and\ \citenamefont {Zaffaroni}}]{Belin:2025hsg}%
  \BibitemOpen
  \bibfield  {author} {\bibinfo {author} {\bibfnamefont {A.}~\bibnamefont {Belin}}, \bibinfo {author} {\bibfnamefont {P.}~\bibnamefont {Singh}}, \bibinfo {author} {\bibfnamefont {R.}~\bibnamefont {Vadala}},\ and\ \bibinfo {author} {\bibfnamefont {A.}~\bibnamefont {Zaffaroni}},\ }\bibfield  {title} {\bibinfo {title} {{Fortuity in ABJM}},\ }\href@noop {} {\  (\bibinfo {year} {2025})},\ \Eprint {https://arxiv.org/abs/2512.04146} {arXiv:2512.04146 [hep-th]} \BibitemShut {NoStop}%
\bibitem [{\citenamefont {Behan}\ and\ \citenamefont {de~Gioia}(2025)}]{Behan:2025hbx}%
  \BibitemOpen
  \bibfield  {author} {\bibinfo {author} {\bibfnamefont {C.}~\bibnamefont {Behan}}\ and\ \bibinfo {author} {\bibfnamefont {L.~P.}\ \bibnamefont {de~Gioia}},\ }\bibfield  {title} {\bibinfo {title} {{Two roads to fortuity in ABJM theory}},\ }\href@noop {} {\  (\bibinfo {year} {2025})},\ \Eprint {https://arxiv.org/abs/2512.23603} {arXiv:2512.23603 [hep-th]} \BibitemShut {NoStop}%
\bibitem [{\citenamefont {Choi}\ \emph {et~al.}(2023)\citenamefont {Choi}, \citenamefont {Kim}, \citenamefont {Lee}, \citenamefont {Lee},\ and\ \citenamefont {Park}}]{Choi:2023znd}%
  \BibitemOpen
  \bibfield  {author} {\bibinfo {author} {\bibfnamefont {S.}~\bibnamefont {Choi}}, \bibinfo {author} {\bibfnamefont {S.}~\bibnamefont {Kim}}, \bibinfo {author} {\bibfnamefont {E.}~\bibnamefont {Lee}}, \bibinfo {author} {\bibfnamefont {S.}~\bibnamefont {Lee}},\ and\ \bibinfo {author} {\bibfnamefont {J.}~\bibnamefont {Park}},\ }\bibfield  {title} {\bibinfo {title} {{Towards quantum black hole microstates}},\ }\href {https://doi.org/10.1007/JHEP11(2023)175} {\bibfield  {journal} {\bibinfo  {journal} {JHEP}\ }\textbf {\bibinfo {volume} {11}},\ \bibinfo {pages} {175}},\ \bibinfo {note} {[Erratum: JHEP 03, 091 (2025)]},\ \Eprint {https://arxiv.org/abs/2304.10155} {arXiv:2304.10155 [hep-th]} \BibitemShut {NoStop}%
\bibitem [{\citenamefont {Choi}\ \emph {et~al.}(2024)\citenamefont {Choi}, \citenamefont {Choi}, \citenamefont {Kim}, \citenamefont {Lee},\ and\ \citenamefont {Lee}}]{Choi:2023vdm}%
  \BibitemOpen
  \bibfield  {author} {\bibinfo {author} {\bibfnamefont {J.}~\bibnamefont {Choi}}, \bibinfo {author} {\bibfnamefont {S.}~\bibnamefont {Choi}}, \bibinfo {author} {\bibfnamefont {S.}~\bibnamefont {Kim}}, \bibinfo {author} {\bibfnamefont {J.}~\bibnamefont {Lee}},\ and\ \bibinfo {author} {\bibfnamefont {S.}~\bibnamefont {Lee}},\ }\bibfield  {title} {\bibinfo {title} {{Finite N black hole cohomologies}},\ }\href {https://doi.org/10.1007/JHEP12(2024)029} {\bibfield  {journal} {\bibinfo  {journal} {JHEP}\ }\textbf {\bibinfo {volume} {12}},\ \bibinfo {pages} {029}},\ \Eprint {https://arxiv.org/abs/2312.16443} {arXiv:2312.16443 [hep-th]} \BibitemShut {NoStop}%
\bibitem [{\citenamefont {de~Mello~Koch}\ \emph {et~al.}(2025)\citenamefont {de~Mello~Koch}, \citenamefont {Kim}, \citenamefont {Kim}, \citenamefont {Lee},\ and\ \citenamefont {Lee}}]{deMelloKoch:2024pcs}%
  \BibitemOpen
  \bibfield  {author} {\bibinfo {author} {\bibfnamefont {R.}~\bibnamefont {de~Mello~Koch}}, \bibinfo {author} {\bibfnamefont {M.}~\bibnamefont {Kim}}, \bibinfo {author} {\bibfnamefont {S.}~\bibnamefont {Kim}}, \bibinfo {author} {\bibfnamefont {J.}~\bibnamefont {Lee}},\ and\ \bibinfo {author} {\bibfnamefont {S.}~\bibnamefont {Lee}},\ }\bibfield  {title} {\bibinfo {title} {{Brane-fused black hole operators}},\ }\href {https://doi.org/10.1007/JHEP07(2025)216} {\bibfield  {journal} {\bibinfo  {journal} {JHEP}\ }\textbf {\bibinfo {volume} {07}},\ \bibinfo {pages} {216}},\ \Eprint {https://arxiv.org/abs/2412.08695} {arXiv:2412.08695 [hep-th]} \BibitemShut {NoStop}%
\bibitem [{\citenamefont {Kim}\ \emph {et~al.}(2003)\citenamefont {Kim}, \citenamefont {Klose},\ and\ \citenamefont {Plefka}}]{Kim:2003rza}%
  \BibitemOpen
  \bibfield  {author} {\bibinfo {author} {\bibfnamefont {N.}~\bibnamefont {Kim}}, \bibinfo {author} {\bibfnamefont {T.}~\bibnamefont {Klose}},\ and\ \bibinfo {author} {\bibfnamefont {J.}~\bibnamefont {Plefka}},\ }\bibfield  {title} {\bibinfo {title} {{Plane wave matrix theory from N=4 superYang-Mills on R x S**3}},\ }\href {https://doi.org/10.1016/j.nuclphysb.2003.08.019} {\bibfield  {journal} {\bibinfo  {journal} {Nucl. Phys. B}\ }\textbf {\bibinfo {volume} {671}},\ \bibinfo {pages} {359} (\bibinfo {year} {2003})},\ \Eprint {https://arxiv.org/abs/hep-th/0306054} {arXiv:hep-th/0306054} \BibitemShut {NoStop}%
\bibitem [{\citenamefont {Lin}\ \emph {et~al.}(2023)\citenamefont {Lin}, \citenamefont {Maldacena}, \citenamefont {Rozenberg},\ and\ \citenamefont {Shan}}]{Lin:2022zxd}%
  \BibitemOpen
  \bibfield  {author} {\bibinfo {author} {\bibfnamefont {H.~W.}\ \bibnamefont {Lin}}, \bibinfo {author} {\bibfnamefont {J.}~\bibnamefont {Maldacena}}, \bibinfo {author} {\bibfnamefont {L.}~\bibnamefont {Rozenberg}},\ and\ \bibinfo {author} {\bibfnamefont {J.}~\bibnamefont {Shan}},\ }\bibfield  {title} {\bibinfo {title} {{Looking at supersymmetric black holes for a very long time}},\ }\href {https://doi.org/10.21468/SciPostPhys.14.5.128} {\bibfield  {journal} {\bibinfo  {journal} {SciPost Phys.}\ }\textbf {\bibinfo {volume} {14}},\ \bibinfo {pages} {128} (\bibinfo {year} {2023})},\ \Eprint {https://arxiv.org/abs/2207.00408} {arXiv:2207.00408 [hep-th]} \BibitemShut {NoStop}%
\bibitem [{\citenamefont {Chen}\ \emph {et~al.}(2025)\citenamefont {Chen}, \citenamefont {Lin},\ and\ \citenamefont {Shenker}}]{Chen:2024oqv}%
  \BibitemOpen
  \bibfield  {author} {\bibinfo {author} {\bibfnamefont {Y.}~\bibnamefont {Chen}}, \bibinfo {author} {\bibfnamefont {H.~W.}\ \bibnamefont {Lin}},\ and\ \bibinfo {author} {\bibfnamefont {S.~H.}\ \bibnamefont {Shenker}},\ }\bibfield  {title} {\bibinfo {title} {{BPS chaos}},\ }\href {https://doi.org/10.21468/SciPostPhys.18.2.072} {\bibfield  {journal} {\bibinfo  {journal} {SciPost Phys.}\ }\textbf {\bibinfo {volume} {18}},\ \bibinfo {pages} {072} (\bibinfo {year} {2025})},\ \Eprint {https://arxiv.org/abs/2407.19387} {arXiv:2407.19387 [hep-th]} \BibitemShut {NoStop}%
\bibitem [{\citenamefont {Chen}\ \emph {et~al.}(2026)\citenamefont {Chen}, \citenamefont {Colin-Ellerin}, \citenamefont {Mamroud},\ and\ \citenamefont {Papadodimas}}]{Chen:2026vml}%
  \BibitemOpen
  \bibfield  {author} {\bibinfo {author} {\bibfnamefont {Y.}~\bibnamefont {Chen}}, \bibinfo {author} {\bibfnamefont {S.}~\bibnamefont {Colin-Ellerin}}, \bibinfo {author} {\bibfnamefont {O.}~\bibnamefont {Mamroud}},\ and\ \bibinfo {author} {\bibfnamefont {K.}~\bibnamefont {Papadodimas}},\ }\bibfield  {title} {\bibinfo {title} {{Chaos of Berry curvature for BPS microstates}},\ }\href@noop {} {\  (\bibinfo {year} {2026})},\ \Eprint {https://arxiv.org/abs/2604.23287} {arXiv:2604.23287 [hep-th]} \BibitemShut {NoStop}%
\bibitem [{\citenamefont {Asano}\ \emph {et~al.}(2015)\citenamefont {Asano}, \citenamefont {Kawai},\ and\ \citenamefont {Yoshida}}]{Asano:2015eha}%
  \BibitemOpen
  \bibfield  {author} {\bibinfo {author} {\bibfnamefont {Y.}~\bibnamefont {Asano}}, \bibinfo {author} {\bibfnamefont {D.}~\bibnamefont {Kawai}},\ and\ \bibinfo {author} {\bibfnamefont {K.}~\bibnamefont {Yoshida}},\ }\bibfield  {title} {\bibinfo {title} {{Chaos in the BMN matrix model}},\ }\href {https://doi.org/10.1007/JHEP06(2015)191} {\bibfield  {journal} {\bibinfo  {journal} {JHEP}\ }\textbf {\bibinfo {volume} {06}},\ \bibinfo {pages} {191}},\ \Eprint {https://arxiv.org/abs/1503.04594} {arXiv:1503.04594 [hep-th]} \BibitemShut {NoStop}%
\bibitem [{\citenamefont {Markovi{\'c}}\ and\ \citenamefont {{\v{C}}ubrovi{\'c}}(2022)}]{Markovic:2022jta}%
  \BibitemOpen
  \bibfield  {author} {\bibinfo {author} {\bibfnamefont {D.}~\bibnamefont {Markovi{\'c}}}\ and\ \bibinfo {author} {\bibfnamefont {M.}~\bibnamefont {{\v{C}}ubrovi{\'c}}},\ }\bibfield  {title} {\bibinfo {title} {{Detecting few-body quantum chaos: out-of-time ordered correlators at saturation}},\ }\href {https://doi.org/10.1007/JHEP05(2022)023} {\bibfield  {journal} {\bibinfo  {journal} {JHEP}\ }\textbf {\bibinfo {volume} {05}},\ \bibinfo {pages} {023}},\ \Eprint {https://arxiv.org/abs/2202.09443} {arXiv:2202.09443 [hep-th]} \BibitemShut {NoStop}%
\bibitem [{\citenamefont {Amore}\ \emph {et~al.}(2025)\citenamefont {Amore}, \citenamefont {Pando~Zayas}, \citenamefont {Pedraza}, \citenamefont {Quiroz},\ and\ \citenamefont {Terrero-Escalante}}]{Amore:2024ihm}%
  \BibitemOpen
  \bibfield  {author} {\bibinfo {author} {\bibfnamefont {P.}~\bibnamefont {Amore}}, \bibinfo {author} {\bibfnamefont {L.~A.}\ \bibnamefont {Pando~Zayas}}, \bibinfo {author} {\bibfnamefont {J.~F.}\ \bibnamefont {Pedraza}}, \bibinfo {author} {\bibfnamefont {N.}~\bibnamefont {Quiroz}},\ and\ \bibinfo {author} {\bibfnamefont {C.~A.}\ \bibnamefont {Terrero-Escalante}},\ }\bibfield  {title} {\bibinfo {title} {{Fuzzy spheres in stringy matrix models: quantifying chaos in a mixed phase space}},\ }\href {https://doi.org/10.1007/JHEP06(2025)031} {\bibfield  {journal} {\bibinfo  {journal} {JHEP}\ }\textbf {\bibinfo {volume} {06}},\ \bibinfo {pages} {031}},\ \Eprint {https://arxiv.org/abs/2407.07259} {arXiv:2407.07259 [hep-th]} \BibitemShut {NoStop}%
\bibitem [{\citenamefont {Maldacena}\ \emph {et~al.}(2016)\citenamefont {Maldacena}, \citenamefont {Shenker},\ and\ \citenamefont {Stanford}}]{Maldacena:2015waa}%
  \BibitemOpen
  \bibfield  {author} {\bibinfo {author} {\bibfnamefont {J.}~\bibnamefont {Maldacena}}, \bibinfo {author} {\bibfnamefont {S.~H.}\ \bibnamefont {Shenker}},\ and\ \bibinfo {author} {\bibfnamefont {D.}~\bibnamefont {Stanford}},\ }\bibfield  {title} {\bibinfo {title} {{A bound on chaos}},\ }\href {https://doi.org/10.1007/JHEP08(2016)106} {\bibfield  {journal} {\bibinfo  {journal} {JHEP}\ }\textbf {\bibinfo {volume} {08}},\ \bibinfo {pages} {106}},\ \Eprint {https://arxiv.org/abs/1503.01409} {arXiv:1503.01409 [hep-th]} \BibitemShut {NoStop}%
\end{thebibliography}%

\end{document}